# FIRST MAGDAS EQUIPMENT IN ECUADOR

E. López[1,2], G. Maeda[3], K. Vicente[1], K. Yumoto[3], N. Vasquez[1],

H. Matsushita[3], A. Shishime[3], C. Vásconez[1]

[1]Observatorio Astronómico de Quito, Escuela Politécnica Nacional, Ecuador;
[2]Space Telescope Science Institute, USA
[3]International Center for Space Weather Science and Education,
Kyushu University, Fukuoka, Japan

E mail (ericsson.lopez@epn.edu.ec, kleber.vicente@epn.edu.ec, maeda@serc.kyushu-u.ac.jp )

*Abstract*. The Magnetic Data Acquisition System (MAGDAS) was installed in the protected Jerusalem Park in Malchingui-Ecuador in October of 2012, under the joint collaboration between Kyushu University of Japan and the Quito Astronomical Observatory of the National Polytechnic School of Ecuador. In this paper, we describe the installation process and present the preliminary data obtained with the MAGDAS equipment. The behavior of the four components, D, H, Z and F allow us to see the importance of having the Ecuador station where the magnetic field has not been systematically measured before, in valuable contribution to study of the equatorial atmosphere.

## Introduction

The MAGDAS Project is a major initiative of the International Center for Space Weather Science and Education (ICSWSE) of Kyushu University of Japan. This project was born in the International Heliophysical Year 2007 (IHY) and today it accounts for around 70 stations deployed mainly in Asia and Africa. In South America there are few stations placed in Peru and Brazil.

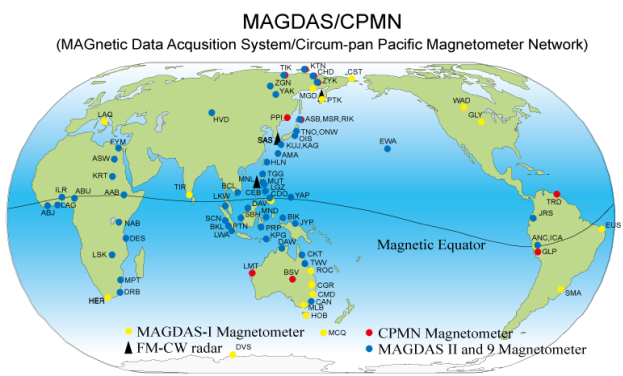

Figure 1: MAGDAS stations worldwide distribution ( ICSWSE, 2013)

Sun variations and its influence on the Earth atmosphere have motivated scientist around the World to investigate the implications on the global climate. These studies involve the study of the upper atmosphere, in particular the ionosphere currents and oscillations and the Earth magnetosphere. The equatorial and low latitude ionosphere usually experience severe instability during the night, especially between the spring and autumn equinoxes, it is thought that this phenomenon is related to the solar wind incidence that produces disturbances resulting in electromagnetic and plasma environment variations of the geo-space.

Beside the African Equatorial space weather (SW) stations it is very important to have further SW stations in the equatorial zone around the globe. Fortunately and thanks to the kind contribution of the United Nations and Space Agencies that organize scientific meetings and workshops on Space Sciences, the Quito Astronomical Observatory (OAQ) of the National Polytechnic School of Ecuador has been motivated and has decided to incorporate inside its scientific interests, the space science studies. This decision has great importance due to the special geographical localization of the Quito Observatory and its facilities just on the equator, making of the observed data important for the scientific community working in understanding the physics behind the phenomena that occur in the equatorial region. These data are useful for modeling the magnetosphere in completion of the extremely valuable data obtained in the Peru station where the Earth magnetic field peaks.

## MAGDAS-Jerusalem Installation Process

The contribution of the United Nation Office for Outer Space Affairs (UNOOSA) with the organization of schools, scientific meetings and Annual Workshops has made possible that Ecuador will be involved in space sciences research in contribution for the study of the equatorial atmosphere. In a couple of these meetings the Director of Quito Observatory has expressed the institutional interest for hosting a MAGDAS equipment in Ecuador, this initiative has been receipted with complacence by Professor K. Yumoto leader scientist of the MAGDAS project of the ICSWSE of Kyushu University. The opportunity came in 2012, year of the realization of the third ISWI, the UN/Ecuador Workshop hold in Quito. The decision for the installation of the MAGDAS magnetometer in the headquarters of the Quito Observatory was taken by Engineer G. Maeda from the Space Environment Research Centre of the Kyushu University, who supervised the installation and did the calibration process of the magnetometer in Ecuador, at the time of occurrence of the UN/Ecuador Workshop in October, 2012.

The best place for the installation of the MAGDAS equipment was chosen, avoiding the electromagnetic noise, regarding soil conditions, access facility, internet connection and other technical requirements. Finally, a place in the Protected Forest of Jerusalem was selected. Jerusalem is located 35 km northward from Quito and it is the neighbor site of our land for the future Astronomical Observatory of Ecuador, located just at the equator (0° 0' 8.67" S latitude).

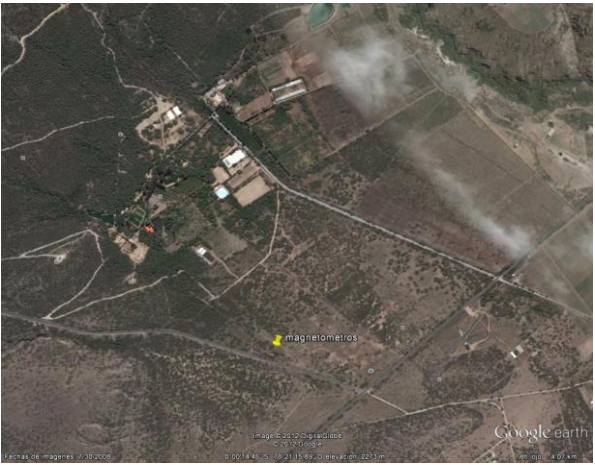

Figure 2: MAGDAS-JERUSALEM site

Regarding the installation of the MAGDAS magnetometer the work was coordinated with Japan prior to the beginning of the UN/Ecuador Workshop. Following the design provided by Eng. Maeda, the work started with the construction of the enclosures for the sensor and for the preamplifier. The Quito Observatory provided the necessary support for the construction of these enclosures, in preparation for the posterior installation of the equipment and its calibration.

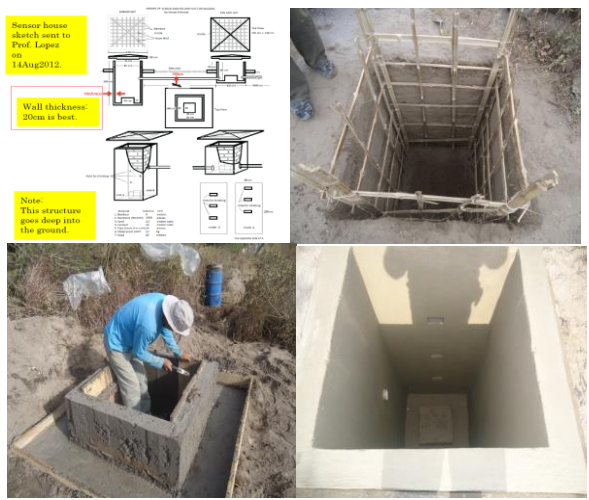

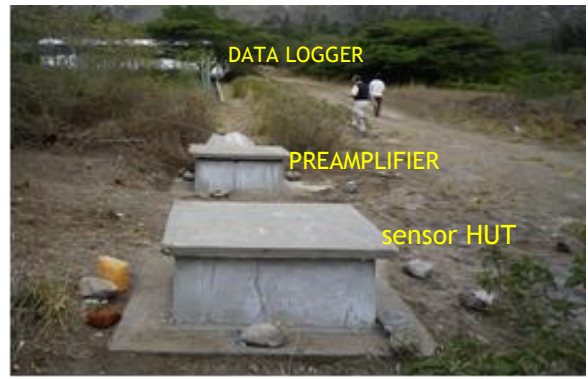

Figure 3: Houses for sensor HUT and preamplifier construction

The construction of the wells continued during the Workshop week, after that, the Quito Observatory technical staff jointly with Eng. G. Maeda and his students of Kyushu University, Japan, installed the sensor in Jerusalem Park. The sensor and preamplifier were installed in wells and covered with 300kg covers to prevent leakage of moisture and the entry of animals.

The elements of the magnetometer were connected with 70 meters of cable buried for its protection in PVC tubes, to avoid the humidity of the soil and to guarantee a good quality signal that enters to the data logger. The local time in the data logger is configured automatically by a GPS antenna mounted on a roof.

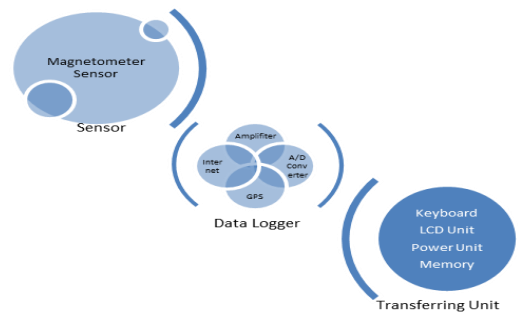

Figure 4: Function Steps of MAGDAS system.

MAGDAS transmits data via the internet to ICSWSE in real time and simultaneously stores the data on an internal data card (Compact Flash). Hence if Internet transmission is disrupted for some reason, the data is always securely saved on this card.

The Internet connection was made with two RF antennas of 900 MHz that link the equipment and the data logger located at the administration Park house. Providing the internet service at this location was a problem solved with the installation of these antennas.

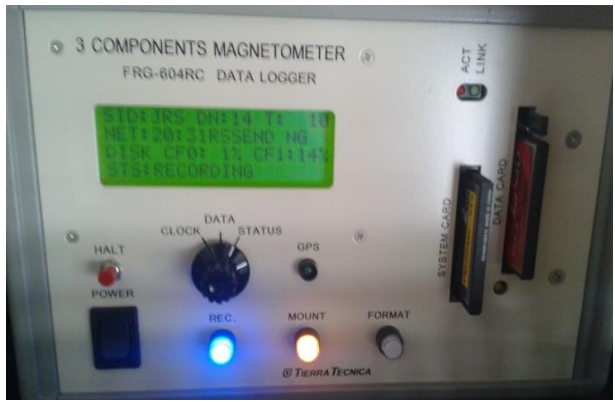

Figure 5: The 3-component magnetometer MAGDAS in Jerusalem, Ecuador.

| MODEL | FRG-604RC |
|---|---|
| MANUFACTURER | TIERRA TECNICA Ltd., Tokyo, Japan |
| PRE AMPLIFIER | A/D converter (magnetic fields, and tilt) |
| | Resolution; 32 bits |
| | Sampling frequency of A/D; 250 Hz |
| | Frequency of data acquisition;1-10Hz |
| | Resolution of magnetic field; 0.01 nT, range; 70,000 nT |
| | Resolution of tilts; 0.1, range; 0.25 degree (900 sec. degree) |
| | Power c.a. 5 W, c.a. 12 V 400 mA |
| | Communication interface; RS-432c |
| | Dimensions, and Weight |
| | Sensor 162mm x 110mm, 4.6 kg |
| | Preamp 160 mm x 260mm x 90mm, 2.9 kg |
| DATA LOGGER | Resolution for recording; 0.1 sec. Degree/ LSB |
| | Power supply Range; from 0 to 25.6 V |
| | Resolution; 0.1 V/ LSB |
| | Data sampled: 1 - 10Hz |
| | GPS clock |
| | Memory SD: File system; FAT16 – max 32GB |
| | Ethernet 10 BASE-T |
| | Operating system: Linux, Kernel 2.4, Distribution: Debian GNU/Linux |
| | Range from AC 90 V to 240 V |
| | Dimension and Weight: 133x210x230mm, 2.6kg |

Table 1: MAGDAS 9 technical features.

The main configuration of the MAGDAS is performed from the Data Logger keyboard and part of this is displayed on the screen of the computer. A backup 12V battery is attached in order to maintain as much as possible the continuity in the data collection process. All the equipment is placed on a metal shelf for its protection and proper functioning.

## Data analysis and results

The geographic coordinates of Jerusalem (JRS) station are Latitude: 0° 0' 8.67" S, Longitude: 78° 21' 24.87" O and Altitude: 2282m. JRS station is 400 meters southward from the Equator.

The MAGDAS magnetometer is completely operative and the system is properly working virtual since its installation. The data are being recorded in a digital database in Quito and currently sharing with the team of the Kyushu University. In recent months the internet system is working well and now we are able to receive and transmit the data remotely.

The personnel of the Quito Astronomical Observatory are handling the data and the first results have been obtained from the first data analysis. The curves shown below are the magnetic field components H and Z, the deviation angle D and the calculated magnetic field for the Jerusalem site.

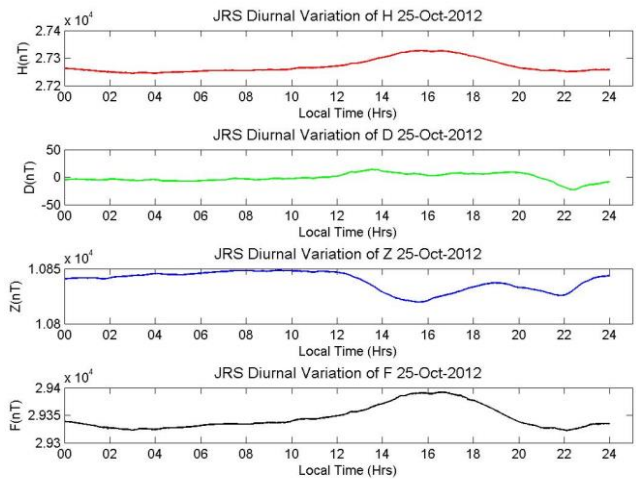

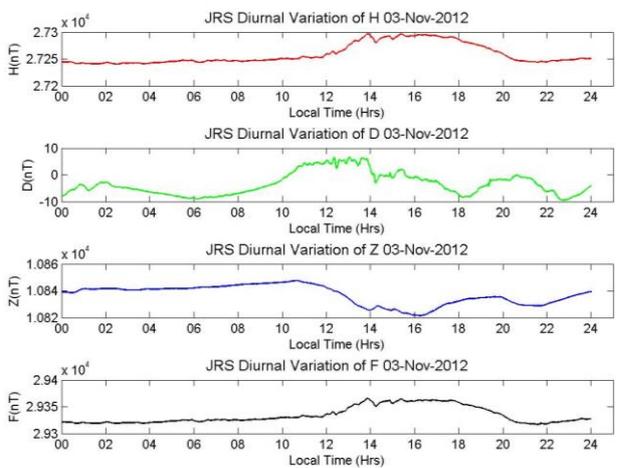

Figure 6: The daily variation of the magnetic field in the MAGDAS Jerusalem Station: H(horizontal component), Z(vertical component), D(deviation angle between H and X component) and F(the calculated geo-magnetic field)

The importance of these data also is in the fact that this is the first time that we are really measuring the magnetic field in Ecuador and we are able now to be involved in the scientific research of interesting phenomena that occur in the equatorial atmosphere. For instances, now we are able to monitor the daily variations of geomagnetic field in our region. In our results we see that the magnetic field is almost constant during the night with values about $2.933 \times 10^4$ (nT) with a remarkable tendency to increase significantly during the daylight. The curves pick around $2.938 \times 10^4$ (nT). These increments of the magnetic field values observed in the data analysis surely are connected with interesting physical processes that take place in the magnetosphere and probably related to the daylight electron density variations of the ionosphere. Currently, we are working on this problem seeking to understand the physical reason for the variation of the observed magnetic field. One explanation that we are reviewing is the occurrence of the equatorial electro jets, but further investigation in this field is necessary.

With the researchers of the Quito Observatory, the students of the National Polytechnic School would be joined for processing a large database of quality values that we expect to have for modeling the magnetic field along the years in association with the Sun activity and its periods.

## Conclusions

Thanks to the kind collaboration of Kyushu University and the dedicated effort of scientists and engineers, Ecuador is measuring the geo-magnetic field in the equatorial region. The MAGDAS equipment is a valuable contribution in order to provide quality data to study the physical phenomena connected with the Sun and its influence in the Earth magnetosphere. The magnetometer is operating properly and the data are being analyzed. We expect soon that the first results of this space science station will appear in scientific journals, with models that help us to improve our understanding about the physical processes that take place in the equatorial atmosphere.

## Acknowledgments

Ecuadorian authors want to express their gratitude to Professor K. Yumoto and Engineer G. Maeda as well as to their team of young students from Kyushu University of Japan, for providing the MAGDAS magnetometer to the Quito Astronomical Observatory. For the huge support received and for their participation in the installation and calibration of the equipment. In the same spirit EL thanks Engineer Kleber Vicente for his dedicated labor and for his professionalism to guarantee the successful installation of the magnetometer and for maintaining its continuous operation. Special thanks to the National Polytechnic School for the support and facilities provided. Finally, many thanks to Gobierno de la Provincia de Pichincha for the kind collaboration to the project, providing the necessary space and required facilities in the Jerusalem Park. E.L. was supported by the National Secretary of Higher Education, Science, Technology and Innovation of Ecuador (Senescyt, fellowship 2011).